# Formation of ultracold LiRb molecules by photoassociation near the Li (2$s$ $^2$S$_{1/2}$) + Rb (5$p$ $^2$P$_{1/2}$) asymptote

Sourav Dutta[1(a)], Daniel S. Elliott[1,2] and Yong P. Chen[1,2]

[1] *Department of Physics, Purdue University, West Lafayette, IN 47907, U.S.A.*
[2] *School of Electrical and Computer Engineering, Purdue University, West Lafayette, IN 47907, U.S.A.*

**Abstract** – We report the production of ultracold $^7$Li$^{85}$Rb molecules by photoassociation (PA) below the Li (2$s$ $^2$S$_{1/2}$) + Rb (5$p$ $^2$P$_{1/2}$) asymptote. We perform PA spectroscopy in a dual-species $^7$Li-$^{85}$Rb magneto-optical trap (MOT) and detect the PA resonances using trap loss spectroscopy. We observe several strong PA resonances corresponding to the last few bound states, assign the lines and derive the long range $C_6$ dispersion coefficients for the Li (2$s$ $^2$S$_{1/2}$) + Rb (5$p$ $^2$P$_{1/2}$) asymptote. We also report an excited-state molecule formation rate ($P_{LiRb}$) of ~$10^7$ s$^{-1}$ and a PA rate coefficient ($K_{PA}$) of ~$4 \times 10^{-11}$ cm$^3$/s, which are both among the highest observed for heteronuclear bi-alkali molecules. These suggest that PA is a promising route for the creation of ultracold ground state LiRb molecules.

**Introduction.** – Ultracold polar molecules have been the cynosure for many applications owing to their large electric dipole moment, which results in a strong, long-range, anisotropic, dipole-dipole interaction between neighboring molecules. This has been exploited in several proposals, for example, on ultracold chemistry [1–3], quantum computing [4], precision measurements [5], and quantum simulations [6]. The first step in the implementation of many of these proposals is to create ultracold polar molecules in their rovibronic ground state where the dipole moment is the largest. This is often difficult, and, several strategies have been implemented to achieve this goal. The two broad approaches are direct cooling of ground state molecules (using buffer gas cooling [7], Stark deceleration [8], laser cooling [9] etc.) or associating ultracold atoms to create ultracold molecules [1,10–12]. The latter approach has been extensively used to create heteronuclear bi-alkali molecules since the constituent alkali atoms are amenable to laser cooling to sub-mK or much lower temperatures, thus facilitating the production of ultracold polar molecules. In particular, two approaches of associating alkali atoms stand out. The first is photoassociation (PA) [10, 11], as implemented, for example, in the case of LiCs [10, 12], and the second is magneto-association (MA) followed by Stimulated Raman Adiabatic Passage (STIRAP), as used in the case of KRb [1, 2]. The second approach, although elegant, is technically more challenging and works only for species with favorable Feshbach resonances [13]. The first approach is simpler and more general, and has been used to produce ultracold polar molecules continuously in their rovibronic ground state using PA followed by spontaneous emission [10, 12, 14]. The population in the ground state is increased either by a suitably chosen PA resonance, as in resonant coupling schemes [15], or by optically pumping population to the ground state [16, 17]. There is considerable interest in furthering these techniques to other heteronuclear combinations either due to the possibility of finding simpler methods for the production of ultracold molecules or due to their higher dipole moments, different quantum statistics, different collisional properties (similar to KRb, reactions between LiRb molecules is exothermic, meaning that it is energetically favorable for LiRb molecules to form Li$_2$ and Rb$_2$ molecules through a reactive collision [18]. Such reactions can be suppressed in an optical lattice, as was done for KRb [19]) etc.

In this article, we report the production of ultracold $^7$Li$^{85}$Rb molecules by PA below the Li (2$s$ $^2$S$_{1/2}$) + Rb (5$p$ $^2$P$_{1/2}$) asymptote, i.e. near the D$_1$ line of atomic Rb (PA near the D$_2$ line is also studied and will be reported in a separate manuscript). The LiRb molecule has recently attracted considerable interest because the rovibronic ground state of the LiRb molecule is predicted to have a relatively high dipole moment of 4.1 Debye [20] (exceeded only by LiCs and NaCs) which makes it a strong candidate for many of the applications mentioned above. That said, LiRb is still one of the least experimentally explored bi-alkali molecule and no ultracold LiRb molecules have been created previously. Accurate spectroscopic information is required for the production of such molecules. While the theoretical potential energy curves [21, 22] have existed since 2000, it is only recently that high resolution spectroscopic studies on hot vapor-phase LiRb have been performed [23–25]. On the ultracold side, Feshbach resonances have recently been reported for the $^7$Li-$^{87}$Rb, $^6$Li-$^{87}$Rb and $^6$Li-$^{85}$Rb systems [26–

[a]E-mail: sourav.dutta.mr@gmail.com

28], although no ultracold molecule formation was reported. In this article, we study the hitherto unexplored ultracold $^7$Li-$^{85}$Rb system and report the production of ultracold $^7$Li-$^{85}$Rb molecules. The $^7$Li$^{85}$Rb molecules are created in excited electronic states (denoted by LiRb*) by photoassociating ultracold $^7$Li and $^{85}$Rb atoms held in a dual-species magneto-optical trap (MOT). This provides accurate spectroscopic information and is the first important step towards creating ultracold ground state LiRb molecules. Contrary to previous expectation [29], we find high PA rates which is promising for ultimately creating ultracold LiRb molecules in their rovibronic ground state.

**Experiment.** – In our experiment, we use a conventional MOT for $^7$Li and a dark MOT for $^{85}$Rb. The $^7$Li MOT is loaded from a Zeeman slower and the typical number ($N_{Li}$) of trapped $^7$Li atoms is around $5\times10^7$ at a density ($n_{Li}$) around $5\times10^9$ cm$^{-3}$. The majority of the trapped $^7$Li atoms are in the upper ($F = 2$) hyperfine level of the $2s\ ^2S_{1/2}$ state. The $^{85}$Rb dark-MOT is loaded from a dispenser and the typical number ($N_{Rb}$) of trapped $^{85}$Rb atoms is around $1\times10^8$ at a density ($n_{Rb}$) around $4\times10^9$ cm$^{-3}$. The majority of the $^{85}$Rb atoms are in the lower ($F = 2$) hyperfine level of the $5s\ ^2S_{1/2}$ state. The trapped atoms collide mainly along the $^7$Li ($2s\ ^2S_{1/2}$, $F = 2$) + $^{85}$Rb ($5s\ ^2S_{1/2}$, $F = 2$) channel. Good spatial overlap of the two MOTs is monitored using a pair of cameras placed orthogonal to each other. The fluorescence from both MOTs is collected using a lens, separated using a dichroic mirror and recorded on two separate photodiodes. More details of our dual species apparatus are described elsewhere [30].

We use a Ti:Sapphire laser for PA. The laser has a linewidth less than 1 MHz, maximum output power of 300 mW and maximum mode-hop-free scan of around 20 GHz. The PA laser beam is collimated to a $1/e^2$ diameter of 0.85 mm, leading to a maximum available peak intensity of about 100 W/cm$^2$. The wavelength of the laser is tuned near 795 nm in order to record the PA spectrum below the Li ($2s\ ^2S_{1/2}$) + Rb ($5p\ ^2P_{1/2}$) asymptote. PA resonances lead to the formation of LiRb* molecules, which either spontaneously decay to electronic ground state LiRb molecules or to free Li and Rb atoms with high kinetic energies. Both mechanisms result in loss of Li and Rb atoms from the MOT leading to a decrease in the MOT fluorescence. The PA resonances, and hence formation of LiRb* molecules, can thus be detected using this PA-induced trap loss spectrum. Trap loss in the Li MOT occurs solely due to LiRb* PA resonances since Li$_2$* PA resonances are weak/absent near 795nm. Trap loss should also be observable in the Rb-MOT but the fluorescence from the Rb dark-MOT is weak and the spectrum is complicated by Rb$_2$* PA resonances. To obtain the LiRb* PA spectrum, we thus record the Li MOT fluorescence as the frequency of the PA laser is scanned (Fig. 1). As an additional check for LiRb* molecules, we have verified that the trap loss features in the Li fluorescence are present only when both Li and Rb atoms are simultaneously trapped.

We note that it is often difficult to observe PA-induced trap loss in heteronuclear systems compared to homonuclear systems. This is because PA in heteronuclear systems occur at a much smaller internuclear separation, e.g. the outer turning point ($R_{out}$) of the LiRb* molecule is much smaller than that of Rb$_2$*, which makes the free-bound transition, at comparable PA laser intensities, much weaker for the former [29]. Nevertheless, we observe several strong trap loss features at relatively low PA intensities, indicating high PA rates.

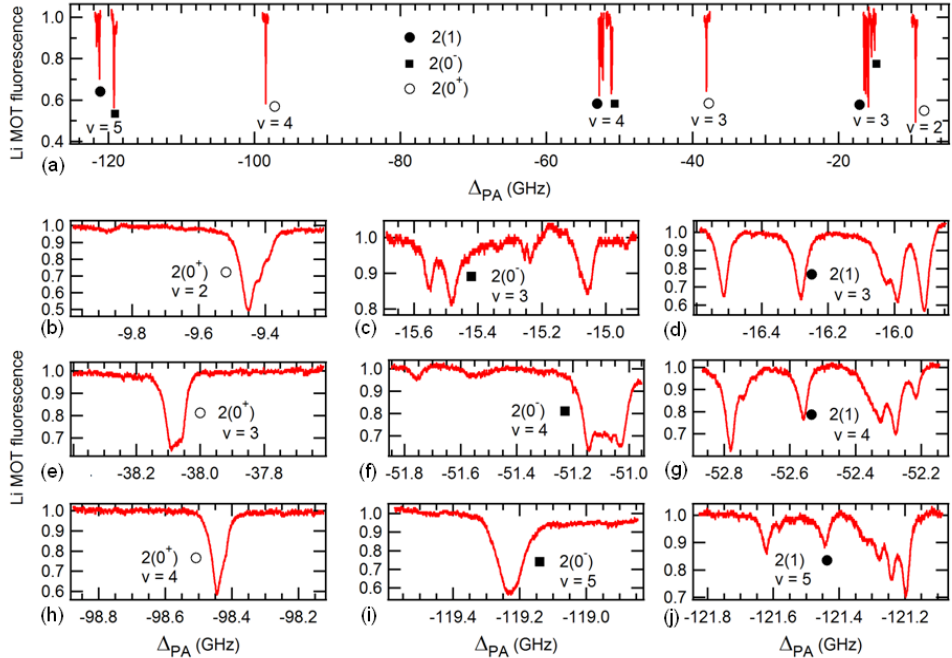

Fig. 1: (Colour on-line) PA spectrum of LiRb* below the Li ($2s\ ^2S_{1/2}$) + Rb ($5p\ ^2P_{1/2}$) asymptote measured using trap loss spectroscopy. (a) Compilation of all strong PA lines. (b-j) Zoom-in view showing the internal structures of the PA lines. The electronic states and vibrational levels (measured from the dissociation limit) are also indicated (see text).

**Results.** – The experimentally observed LiRb* PA resonances below the Li ($2s\ ^2S_{1/2}$) + Rb ($5p\ ^2P_{1/2}$) asymptote are shown in Figure 1. The spectrum in Figure 1a is obtained by stitching together several short PA scans shown in Figures 1b-j. The detuning $\Delta_{PA}$ of the PA laser is measured with respect to the frequency $\nu_{res}$ (= 377108.946 GHz) of the Rb ($5s\ ^2S_{1/2}$, $F = 2$) → Rb ($5p\ ^2P_{1/2}$, $F' = 2$) transition i.e. $\Delta_{PA} = \nu_{PA} - \nu_{res}$, where $\nu_{PA}$ is the frequency of the PA laser. The detuning $\Delta_{PA}$ is thus a measure of the binding energies ($E_B$) of the respective ro-vibrational levels. The spectrum is obtained by scanning the frequency of the PA laser at around 1.9 MHz/s and with a peak intensity ~100 W/cm$^2$. The absolute accuracy in the measurement of PA line frequency is 100 MHz, determined primarily by the accuracy (60 MHz) of the wavelength-meter used for these measurements. The relative accuracy of the frequency measurement and the frequency resolution are however much better. The full width at half maximum (FWHM) linewidth of the narrowest lines are found to be ~33 MHz, which is reasonable given the natural linewidth (~12 MHz) and the thermal broadening (~21 MHz) due to the temperature (1 mK) of a typical Li MOT.

The assignment of the PA lines is facilitated by the availability of theoretical values for the long range $C_6$ dispersion coefficients [31–33]. Figure 2a shows the first few electronic states of the LiRb molecule for small internuclear separations [22]. In the present PA experiment, we probe levels close to the Li ($2s\ ^2S_{1/2}$) + Rb ($5p\ ^2P_{1/2}$) asymptote for which the internuclear separation is larger and it is more appropriate to use the Hund's case (c) notation. As shown in Figure 2b, there are three, almost identical, electronic states asymptotic to the Li ($2s\ ^2S_{1/2}$) + Rb ($5p\ ^2P_{1/2}$) asymptote. These are labeled $n(\Omega^\sigma)$ [10, 11], where $\Omega$ is the projection of the total electronic angular momentum on the internuclear axis, $\sigma = -/+$ (only for $\Omega = 0$ states) depending on whether or not the electronic wave function changes sign upon reflection at any plane containing the internuclear axis, and $n$ is a number denoting the $n^{th}$ electronic state of a particular $\Omega^\sigma$. The potential energy curves of the three states, $2(0^+)$, $2(0^-)$ and $2(1)$, are almost identical since at such large internuclear separations the electronic potentials are almost entirely determined by the $C_6$ coefficients which are very similar for the three electronic states. Experimentally, however, it is possible to determine three distinct, although similar, $C_6$ coefficients for these states. We note that there are five other electronic states, $3(0^+)$, $3(0^-)$, $3(1)$, $4(1)$ and $1(2)$, converging to the Rb $D_2$ asymptote but their transition strengths are expected to be very weak in the frequency range (below the Rb $D_1$ asymptote) probed here. We thus consider only the $2(0^+)$, $2(0^-)$ and $2(1)$ states in this report.

We group the observed PA lines based on their structure and assign them to one of the three potentials (table 1). For example, the $\Omega = 1$ state is expected to have hyperfine structure and we thus assign the lines with multiple structures to the $2(1)$ state. The $\Omega = 0$ states, on the other hand, are expected to have no hyperfine structure (in the first order). Note that the $2(0^-)$ has internal structure for the least bound states but the structure decreases (and goes away) for more deeply bound levels. We believe that the structure in the loosely bound levels of the $2(0^-)$ state is due to mixing with the nearby $2(1)$ state which lies very close to the $2(0^-)$ state. For deeply bound levels, the $2(0^-)$ and $2(1)$ states are well-separated, the mixing become negligible and the $2(0^-)$ state has negligible internal structure as expected. The internal structure (which is attributed to the hyperfine structure) in the $2(1)$ state persists even for deeply bound levels, as expected. We distinguish between the $2(0^+)$ and the $2(0^-)$ states based on the fact that the $2(0^-)$ state is expected to lie closer to the $2(1)$ state [34] but also note that their assignment may need to be interchanged when more spectroscopic data become available. We assign the vibrational level $v$ ($v$ is measured from the asymptote such that $v = 1$ is the least bound level) for each electronic state such that it is consistent with the LeRoy-Bernstein (LRB) formula [35] as discussed later. We note that we observe only one rotational ($J$) level for each vibrational level, which, as discussed below, is somewhat unexpected.

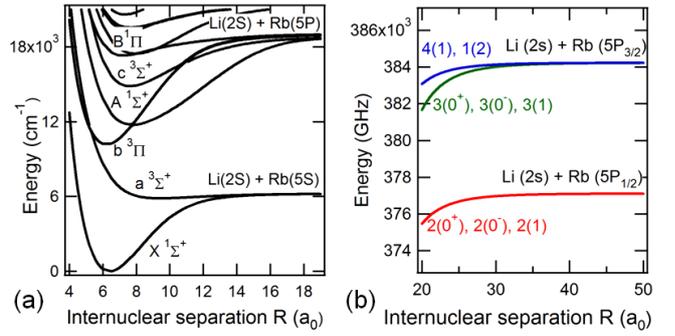

Fig. 2: (Colour on-line) (a) Theoretical potential energy curves of LiRb for small internuclear separations [22]. (b) Theoretical potential energy curves at large internuclear separations determined solely by the $C_6$ coefficients [31]. Only these curves are relevant to the presented PA measurements.

The angular momentum selection rules for the electric dipole allowed transitions are: $\Delta\Omega = 0, \pm1$, $\Delta J = 0, \pm1$, $0^+ \leftrightarrow 0^+$, $0^- \leftrightarrow 0^-$, $J \geq \Omega$, $\Delta J \neq 0$ if $\Omega = 0$ for both states and $J = 0 \rightarrow J = 0$ transitions are not allowed. There are three possible electronic potentials, $1(0^+)$, $1(0^-)$ and $1(1)$, along which collisions between ground state Li and Rb atoms can occur. These collisions are primarily $s$-wave ($l = 0$) collisions since the centrifugal barrier for all higher partial waves ($l > 0$) is high enough that the ultracold atoms do not have sufficient kinetic energy to overcome the barrier and come close enough for PA to occur. For example, we estimate the height of the $p$-wave ($l = 1$) centrifugal barrier at ~100 $a_0$ to be ~1.9 mK using theoretical value of $C_6$ (= 2500 a.u.) for the Li ($2s\ ^2S_{1/2}$) + Rb ($5s\ ^2S_{1/2}$) asymptote [32]. This barrier is higher than the typical temperature (~1 mK) of the Li MOT (the Rb MOT is much colder). Note that the observed PA levels have outer turning points $R_{out}$ in the range 30 – 44.6 $a_0$ (see below); so, in accordance with the Born-Oppenheimer approximation for the free-bound transition, the internuclear distance between the colliding atoms must also be in the same 30 – 44.6 $a_0$ range for PA to occur. The $p$-wave centrifugal barrier at ~100 $a_0$ inhibits atoms from getting close enough where PA occurs, restricting collisions primarily to $l = 0$ ($s$-wave). Thus, in accordance with the angular momentum selection rules mentioned above, the photoassociation of colliding ($s$-wave) Li and Rb atoms can lead to the formation of LiRb* molecules in the following states: $\Omega = 0^+$ ($J = 0, 1, 2$), $\Omega = 0^-$ ($J = 0, 1, 2$) and $\Omega = 1$ ($J = 1, 2$). In light of the preceding discussion, it is expected that up to three rotational levels would be observed but we observe only one rotational level for each vibrational level. For example, the $J = 0$ line for the $2(0^+)$, $v = 2$ level is expected to be at -9.73 GHz (based on the estimated rotational constant of ~140 MHz) but no such line is observed in Figure 1b. The absence of rotational structure is unexplained at present and can make our rotational assignment somewhat ambiguous. We assign $J = 1$ for all observed levels, against $J = 0$ or $J = 2$, taking into

Sourav Dutta *et al.*

Table 1: The values of $-\Delta_{PA}$ (in GHz) for which PA resonances are observed. Also included is the assignment of the electronic states and the vibrational quantum numbers $v$ (measured from dissociation limit). For vibrational levels with multiple internal structures, the frequency of line closest to the center of the spectrum is reported (the internal structure leads to a maximum uncertainty of ~250 MHz in position of the lines reported). We did not observe the next more deeply bound vibrational state probably due weak Franck-Condon overlap with the scattering state. The least bound states (bound by less than 8 GHz) could not be observed because the PA laser, with frequency close to $D_1$ transition, strongly perturbed the operation of the Rb MOT.

| State | $v = 2$ | $v = 3$ | $v = 4$ | $v = 5$ |
|---|---|---|---|---|
| $2(0^+)$ | 9.45 | 38.08 | 98.44 | |
| $2(0^-)$ | | 15.48 | 51.08 | 119.23 |
| $2(1)$ | | 16.28 | 52.56 | 121.44 |

consideration the fact that there are at least twice as many allowed PA transitions from the $1(0^+)$, $1(0^-)$ and $1(1)$ collision channels that could lead to the formation of $J = 1$ molecules in the excited states. We have recently confirmed the assignment of $J = 1$ using Raman-type two-photon photoassociation, the details of which are beyond the scope of this article. The assignments of all electronic and vibration states of the observed PA lines are reported in table 1.

**$C_6$ coefficients.** – As stated earlier, the electronic potentials at such large internuclear separations ($R$) are almost entirely determined by the $C_6$ coefficients, which can be extracted using the LeRoy-Bernstein (LRB) formula [35]:

$$D - E_v = A_6(v - v_D)^3 \quad (1)$$

where $v$ is the vibrational quantum number (measured from the dissociation limit), $v_D$ is vibrational quantum number at dissociation such that $0 < v_D < 1$, $-(D - E_v)$ is the (negative) binding energy ($E_B$), $E_v$ is the energy of the $v^{th}$ vibrational level, $D$ is the dissociation energy and $A_6 = 16\sqrt{2}\pi^3\hbar^3/[B(2/3,1/2)\mu^{3/2}C_6^{1/2}]$. Here $\mu$ (= 6.48 atomic mass unit) is the reduced mass of $^7$Li$^{85}$Rb and $B$ is the Beta function [$B(2/3, 1/2) = 2.5871$]. The experimentally measured quantity is $\Delta_{PA}$ which is related to the binding energy according to the relation $(D - E_v) - E_{rot} = -h\Delta_{PA}$, where $E_{rot} = B_v[J(J+1) - \Omega^2]$ is the rotational energy. We initially ignore the rotational energy and plot $(-h\Delta_{PA})^{1/3} \approx (D - E_v)^{1/3}$ against $v$ and, from a fit to Eq. (1), derive the values of $v_D$ and $A_6$ (and hence $C_6$). The fits for different electronic states are shown in Fig. 3.

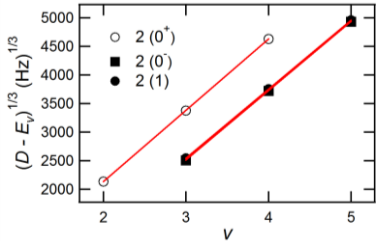

Fig. 3: (Colour on-line) The fit of experimentally observed PA frequencies to the LRB formula as described in the text. Note that the data points for the $2(0^-)$ and $2(1)$ states lie almost on top of each other. Also note that the slopes of all the curves are similar indicating that the $C_6$ coefficients for all three electronic states are similar.

Table 2: The values of $C_6$ coefficients (in atomic units = Hartree/$a_0^6$ ≈ 0.957342 × 10$^{-79}$ Jm$^6$) measured experimentally in this work, and a comparison with three different theoretical predictions. The superscripts and subscripts on the experimental values indicate the uncertainties that arise when $J$ is changed by ±1. For PA resonances with internal structures, the uncertainty in the determination of the center frequency leads to some uncertainty in the determination of the $C_6$ coefficients but the $C_6$ values are still within the quoted uncertainties. The experimentally determined values of $v_D$ are also included (typical uncertainty in $v_D$ is $^{-0.04}_{+0.02}$).

| State | This work | | Ref. 31 | Ref. 32 | Ref. 33 |
|---|---|---|---|---|---|
| | $C_6$ | $v_D$ | $C_6$ | $C_6$ | $C_6$ |
| $2(0^+)$ | $11335^{+600}_{-300}$ | 0.288 | 13900 | 16072 | $^{1,3}\Sigma^+$: 26744 |
| $2(0^-)$ | $13470^{+540}_{-270}$ | 0.927 | 13900 | 16071 | |
| $2(1)$ | $13730^{+535}_{-135}$ | 0.893 | 13900 | 16071 | $^{1,3}\Pi$: 9431 |

In order to account for the rotational energy $E_{rot}$, the following procedure is used. The electronic potential is written in the form $V(R) = -C_6/R^6$, with the $C_6$ coefficients derived above. At the outer turning point $R_{out}$ of the $v^{th}$ vibrational level $V(R_{out}) = -(D - E_v) \approx h\Delta_{PA}$, which implies that $R_{out} \approx (-C_6/h\Delta_{PA})^{1/6}$ ($R_{out}$ lies in the range 30 – 44.6 $a_0$ depending on $v$). The rotational constant is then approximated as $B_v \approx \hbar^2/2\mu R_{out}^2$ and these values lie in the range 140 – 310 MHz depending on $v$. The value of $E_{rot}$ is then calculated and the correction due to $E_{rot}$ is taken into account by plotting $(-h\Delta_{PA} + E_{rot})^{1/3}$ against $v$, and deriving a new set of the values of $v_D$ and $A_6$ (and hence $C_6$). We find that the new set of values of $v_D$ and $C_6$ are very close to the initial values and the values are reported in table 2. We also found that the uncertainties (±1) in the assignment of $J$, as mentioned above, has only minor effects on the values of $v_D$ and $C_6$ extracted and they lie within the quoted uncertainties. Table 2 also includes a comparison of the experimentally determined $C_6$ coefficients and different theoretically calculated values. We find good agreement (within 20%) with the calculated values reported in Ref. [31], considering that perturbations are abundant in molecular systems, that the atomic polarizabilities are difficult to calculate accurately and that the calculations neglected hyperfine interactions, non-adiabatic corrections etc. The agreement with Ref. [33] is not good because their calculations neglected the fine structure of atoms (that distinguishes the $D_1$ and $D_2$ asymptotes).

**Photoassociation rate.** – We note that we observe high LiRb* molecule formation rate ($P_{LiRb}$). $P_{LiRb}$ is determined from the time evolution of the Li MOT fluorescence after the on-resonance PA beam is suddenly applied to the MOT. On applying the PA beam, the Li atom number starts to decrease at a rate $dN_{Li}/dt$ due to the formation of excited LiRb* molecules, which are lost from the MOT. Since one Li atom is lost per LiRb* molecule formed, the rate $-dN_{Li}/dt$ is equal to the molecule formation rate ($P_{LiRb}$). For the $v = 3$ level of the $2(0^+)$ state we observe $P_{LiRb}$ of around $10^7$ s$^{-1}$, which is among the highest observed for heteronuclear bi-alkali molecules. This could lead to formation of a substantial number of electronic ground state LiRb molecules even if only a small fraction of these LiRb* molecules spontaneously decay to the electronic ground state.

Indeed, we have recently detected LiRb molecules in the electronic ground state (although not in the vibrational ground state) using Resonance Enhanced Multi-Photon Ionization (REMPI), the details of which will be discussed elsewhere.

For the same level, we also determine the PA rate coefficient $K_{PA}$ which, for a small Li MOT immersed in a larger Rb MOT, is given by $K_{PA} = P_{LiRb}/(n_{Rb}N_{Li})$. For the measurement of $K_{PA}$, we thus intentionally load a much smaller Li MOT inside a large Rb MOT (such that the Li MOT is completely immersed in the Rb MOT) and derive $K_{PA}$ to be $\sim 4\times 10^{-11}$ cm$^3$/s at a relatively low PA intensity of $\sim$ 90 W/cm$^2$ ($K_{PA}$ is not saturated at this intensity and the saturated $K_{PA}$ value is expected to be higher). The major uncertainty in the measurement of $K_{PA}$ comes from the uncertainty in the measurement of the Rb MOT density $n_{Rb}$ and possible inhomogeneities in $n_{Rb}$ (note that the uncertainty in the measurement of Li MOT atom number $N_{Li}$ cancels out). We conservatively estimate the reported value of $K_{PA}$ to be accurate within a factor of two. The observed PA rate ($K_{PA}$) for LiRb is the highest among all heteronuclear bi-alkali molecules (only LiCs is reported to have a comparable value of $K_{PA}$ of $\sim 10^{-11}$ cm$^3$/s but the measurements were performed using REMPI detection with which only an order of magnitude estimate is possible [36]). This suggests that PA is very efficient in LiRb and hence is an attractive option for the production of electronic ground state molecules.

**Conclusion.** – In conclusion, we report the production of ultracold LiRb* molecules by photoassociation below the Li (2s $^2$S$_{1/2}$) + Rb (5p $^2$P$_{1/2}$) asymptote. We assign the PA spectra and determine the $C_6$ dispersion coefficients for the first time. We also find high PA rates and report a LiRb* molecule formation rate of around $10^7$ s$^{-1}$. We speculate that PA followed by spontaneous emission could potentially lead to efficient production of ground state LiRb molecules.

\*\*\*

We thank John Lorenz and Adeel Altaf for assistance with the experimental apparatus. Financial support from National Science Foundation (CCF-0829918) and Army Research Office (W911NF-10-1-0243) during the early stages of this work is acknowledged. S. Dutta also acknowledges support from Purdue University in the form of the Bilsland Dissertation Fellowship.

*Note added in proofs*: We recently observed high PA rates and unitarity-limited saturation of PA rates for PA near the Li (2s $^2$S$_{1/2}$)+Rb (5p $^2$P$_{3/2}$) asymptote [37].

REFERENCES

[1] NI K. K., OSPELKAUS S., DE MIRANDA M. H. G., PE'ER A., NEYENHUIS B., ZIRBEL J. J., KOTOCHIGOVA S., JULIENNE P. S., JIN D. S. and YE J., *Science*, **322** (2008) 231–235.
[2] OSPELKAUS S., NI K. K., WANG D., DE MIRANDA M. H. G., NEYENHUIS B., QUÉMÉNER G., JULIENNE P. S., BOHN J. L., JIN D. S. and YE J., *Science*, **327** (2010) 853–857.
[3] NI K. K., OSPELKAUS S., WANG D., QUÉMÉNER G., NEYENHUIS B., DE MIRANDA M. H. G., BOHN J. L., YE J. and JIN D. S., *Nature*, **464** (2010) 1324–1328.
[4] DEMILLE D., *Phys. Rev. Lett.*, **88** (2002) 067901.
[5] HUDSON E. R., LEWANDOWSKI H. J., SAWYER B. C. and YE J., *Phys. Rev. Lett.*, **96** (2006) 143004.
[6] BARANOV M. A., DALMONTE M., PUPILLO G. and ZOLLER P., *Chem. Rev.*, **112** (2012) 5012–5061.
[7] HUTZLER N. R., LU H. and DOYLE J. M., *Chem. Rev.*, **112** (2012) 4803–4827.
[8] VAN DE MEERAKKER S. Y. T., BETHLEM H. L. and MEIJER G., *Nature Phys.*, **4** (2008) 595–602.
[9] SHUMAN E. S., BARRY J. F. and DEMILLE D., *Nature*, **467** (2010) 820–823.
[10] ULMANIS J., DEIGLMAYR J., REPP M., WESTER R. and WEIDEMÜLLER M., *Chem. Rev.*, **112** (2012) 4890–4927.
[11] JONES K. M., TIESINGA E., LETT P. D. and JULIENNE P. S., *Rev. Mod. Phys.*, **78** (2006) 483–535.
[12] DEIGLMAYR J., GROCHOLA A., REPP M., MÖRTLBAUER K., GLÜCK C., LANGE J., DULIEU O., WESTER R. and WEIDEMÜLLER M., *Phys. Rev. Lett.*, **101** (2008) 133004.
[13] CHIN C., GRIMM R., JULIENNE P. and TIESINGA E., *Rev. Mod. Phys.*, **82** (2010) 1225–1286.
[14] ZABAWA P., WAKIM A., HARUZA M. and BIGELOW N. P., *Phys. Rev. A*, **84** (2011) 061401.
[15] STWALLEY W. C., BANERJEE J., BELLOS M., CAROLLO R., RECORE M. and MASTROIANNI M., *J. Phys. Chem. A*, **114** (2010) 81–86.
[16] VITEAU M., CHOTIA A., ALLEGRINI M., BOULOUFA N., DULIEU O., COMPARAT D. and PILLET P., *Science*, **321** (2008) 232–234.
[17] SAGE J. M., SAINIS S., BERGEMAN T. and DEMILLE D., *Phys. Rev. Lett.*, **94** (2005) 203001.
[18] ŻUCHOWSKI P. S. and HUTSON J. M., *Phys. Rev. A*, **81** (2010) 060703(R).
[19] CHOTIA A., NEYENHUIS B., MOSES S. A., YAN B., COVEY J. P., FOSS-FEIG M., REY A. M., JIN D. S. and YE J., *Phys. Rev. Lett.*, **108** (2012) 080405.
[20] AYMAR M. and DULIEU O., *J. Chem. Phys.*, **122** (2005) 204302.
[21] KOREK M., ALLOUCHE A. R., KOBEISSI M., CHAALAN A., DAGHER M., FAKHERDDIN K. and AUBERT-FRÉCON M., *Chem. Phys.*, **256** (2000) 1–6.
[22] KOREK M., YOUNES G. and AL-SHAWA S., *J. Mol. Struct.: THEOCHEM*, **899** (2009) 25–31.
[23] IVANOVA M., STEIN A., PASHOV A., KNÖCKEL H. and TIEMANN E., *J. Chem. Phys.*, **134** (2011) 024321.
[24] DUTTA S., ALTAF A., ELLIOTT D. S. and CHEN Y. P., *Chem. Phys. Lett.*, **511** (2011) 7–11.
[25] IVANOVA M., STEIN A., PASHOV A., KNÖCKEL H. and TIEMANN E., *J. Chem. Phys.*, **138** (2013) 094315.
[26] MARZOK C., DEH B., ZIMMERMANN C., COURTEILLE PH. W., TIEMANN E., VANNE Y. V. and SAENZ A., *Phys. Rev. A*, **79** (2009) 012717.
[27] LI Z., SINGH S., TSCHERBUL T. V. and MADISON K. W., *Phys. Rev. A*, **78** (2008) 022710.
[28] DEH B., GUNTON W., KLAPPAUF B. G., LI Z., SEMCZUK M., VAN DONGEN J. and MADISON K. W., *Phys. Rev. A*, **82** (2010) 020701.
[29] WANG H. and STWALLEY W. C., *J. Chem. Phys.*, **108** (1998) 5767–5771.
[30] DUTTA S., ALTAF A., LORENZ J., ELLIOTT D. S. and CHEN Y. P., *Interspecies collision-induced losses in a dual species $^7$Li-$^{85}$Rb magneto-optical trap*, *J. Phys. B: At. Mol. Opt. Phys.* **47** (2014) 105301.
[31] MOVRE M., BEUC R., *Phys. Rev. A*, **31** (1985) 2957–2967.
[32] BUSSERY B., ACHKAR Y. and AUBERT-FRÉCON M., *Chem. Phys.*, **116** (1987) 319–338.
[33] MARINESCU M. and SADEGHPOUR H. R., *Phys. Rev. A*, **59** (1999) 390–404.
[34] STWALLEY W. C., BELLOS M., CAROLLO R., BANERJEE J. and BERMUDEZ M., *Mol. Phys.*, **110** (2012) 1739–1755.
[35] LEROY R. J. and BERNSTEIN R. B., *J. Chem. Phys.*, **52** (1970) 3869–3879.

[36] DEIGLMAYR J., PELLEGRINI P., GROCHOLA A., REPP M., CÔTÉ R., DULIEU O., WESTER R. and WEIDEMÜLLER M., *New J. Phys.*, **11** (2009) 055034.

[37] DUTTA S., LORENZ J., ALTAF A., ELLIOTT D. S. and CHEN Y. P., *Photoassociation of ultracold LiRb\* molecules: observation of high efficiency and unitarity-limited rate saturation*, *Phys. Rev. A*, **89** (2014) 020702(R).